# Introduction to tensorial resistivity probability tomography

**Paolo Mauriello**[1] and **Domenico Patella**[2]

[1]*Department of Science and Technology for Environment and Territory, University of Molise, Campobasso, Italy*
*(E-mail: mauriello@unimol.it)*

[2]*Department of Physical Sciences, University Federico II, Naples, Italy*
*(E-mail: patella@na.infn.it)*

**ABSTRACT**

The probability tomography approach developed for the scalar resistivity method is here extended to the 2D tensorial apparent resistivity acquisition mode. The rotational invariant derived from the trace of the apparent resistivity tensor is considered, since it gives on the datum plane anomalies confined above the buried objects. Firstly, a departure function is introduced as the difference between the tensorial invariant measured over the real structure and that computed for a reference uniform structure. Secondly, a resistivity anomaly occurrence probability (RAOP) function is defined as a normalised crosscorrelation involving the experimental departure function and a scanning function derived analytically using the Frechet derivative of the electric potential for the reference uniform structure. The RAOP function can be calculated in each cell of a 3D grid filling the investigated volume, and the resulting values can then be contoured in order to obtain the 3D tomographic image. Each non-vanishing value of the RAOP function is interpreted as the probability which a resistivity departure from the reference resistivity obtain in a cell as responsible of the observed tensorial apparent resistivity dataset on the datum plane. A synthetic case shows that the highest RAOP values correctly indicate the position of the buried objects and a very high spacial resolution can be obtained even for adjacent objects with opposite resistivity contrasts with respect to the resistivity of the hosting matrix. Finally, an experimental field case dedicated to an archaeological application of the resistivity tensor method is presented as a proof of the high resolution power of the probability tomography imaging, even when the data are collected in noisy open field conditions.

## INTRODUCTION

High-resolution geoelectrical data acquisition tools are now routinely applied to solve many practical problems in environmental and civil engineering, cultural heritage and non-destructive testing. In order to elicit the highest amount of information from large datasets, more target-oriented 3D imaging techniques are demanded. It should be stressed that, in near-surface geophysics, the relevant information is generally limited to the identification and spatial collocation of the buried physical sources of the anomalies detected on the soil, since nature and quality of the materials to be identified are normally known in advance.

Among the various approaches to geophysical data imaging, probability tomography is gaining increasing attention as to objectivity and reliability. The principles of the 3D probability tomography in applied geophysics were established for the self-potential method (Patella, 1997 a,b), then exported to the geoelectric method based on the conventional scalar apparent resistivity concept (Mauriello *et al*., 1998; Mauriello and Patella, 1999), to the electromagnetic induction methods (Mauriello and Patella, 1999, 2000), and to the gravity (Mauriello and Patella, 2001 a,b) and magnetic methods (Mauriello and Patella, 2005). The resolution power of the probability tomography imaging, using either one or a combination of geophysical investigation methods has been largely documented in recent years in earth science (Di Maio *et al*., 1998; Iuliano *et al*., 2001, 2002 a,b; Lapenna *et al*., 2000; Mauriello *et al*., 2004) and in archaeological and cultural heritage (Cammarano *et al*., 2000 a,b; Patella and Mauriello, 1999) applications.

The purpose of this paper is to further extend the 3D probability tomography imaging approach to the geoelectrical method based on the less commonly utilised but still very promising apparent resistivity tensor concept (Bibby, 1977).

## APPARENT RESISTIVITY TENSOR ANALYSIS

### *The concept of apparent resistivity tensor*

Consider a generic 3D resistivity structure buried below a flat free surface. Assume that a resistivity survey has been performed inside a rectangular area *S* using distinct bipole current sources. Following Bibby (1977), for the generic *i*-th bipole source the law

$$\mathbf{E}_i = \rho^{(a)} \mathbf{J}_i \tag{1}$$

can be introduced to relate the measured electrical field vector $\mathbf{E}_i$ to the current density vector $\mathbf{J}_i$ for a uniform half-space, using the concept of apparent resistivity tensor $\rho^{(a)}$.

Using two bipoles ($i=1,2$), $\rho^{(a)}$ is given as (Bibby and Hohmann, 1993)

$$\rho^{(a)} = \begin{pmatrix} \rho_{11}^{(a)} & \rho_{12}^{(a)} \\ \rho_{21}^{(a)} & \rho_{22}^{(a)} \end{pmatrix} =$$

$$\frac{\begin{pmatrix} E_{1x}J_{2y} - E_{2x}J_{1y} & E_{2x}J_{1x} - E_{1x}J_{2x} \\ E_{1y}J_{2y} - E_{2y}J_{1y} & E_{2y}J_{1x} - E_{1y}J_{2x} \end{pmatrix}}{J_{1x}J_{2y} - J_{2x}J_{1y}}. \tag{2}$$

Using this form one can define rotational invariants, which are independent of the direction of the electrical field and the individual current source bipoles. We consider the invariant $P$ related to the trace of $\rho^{(a)}$ (Bibby, 1986), which has the important property of providing anomalies closely confined about the sources. $P$ is given as

$$P = \frac{1}{2}\left[\rho_{11}^{(a)} + \rho_{22}^{(a)}\right] = \frac{1}{2} \frac{E_{1x}J_{2y} - E_{2x}J_{1y} + E_{2y}J_{1x} - E_{1y}J_{2x}}{J_{1x}J_{2y} - J_{2x}J_{1y}}. \tag{3}$$

Take now a reference coordinate system ($x,y,z$) with $S$ on the $xy$-plane and the $z$-axis positive downwards. Putting with $\mathbf{r}_{A_i} \equiv (x_{A_i}, y_{A_i}, 0)$ and $\mathbf{r}_{B_i} \equiv (x_{B_i}, y_{B_i}, 0)$ the coordinates of two fixed current electrodes pairs $A_i$ (positive) and $B_i$ (negative) ($i=1,2$), and with $I_i$ ($i=1,2$) the intensity of the energizing current through the $i$-th current bipole $A_iB_i$, the $J_{ix}$ and $J_{iy}$ ($i=1,2$) terms in eq.3 at the generic variable point $\mathbf{r} \equiv (x,y,0)$, where the electrical field components are measured, are expressed as

$$J_{ix} = \frac{I_i}{2\pi}\left(\frac{\mathbf{r} - \mathbf{r}_{A_i}}{|\mathbf{r} - \mathbf{r}_{A_i}|^3} - \frac{\mathbf{r} - \mathbf{r}_{B_i}}{|\mathbf{r} - \mathbf{r}_{B_i}|^3}\right) \cdot \mathbf{i}, \tag{4a}$$

$$J_{iy} = \frac{I_i}{2\pi}\left(\frac{\mathbf{r} - \mathbf{r}_{A_i}}{|\mathbf{r} - \mathbf{r}_{A_i}|^3} - \frac{\mathbf{r} - \mathbf{r}_{B_i}}{|\mathbf{r} - \mathbf{r}_{B_i}|^3}\right) \cdot \mathbf{j}. \tag{4b}$$

where $\mathbf{i}$ and $\mathbf{j}$ are the unit vectors defining the $x$-axis and $y$-axis, respectively.

*The tensor invariant departure concept*

Assume that the subsoil is made of $Q$ elementary cells with constant volume $\Delta V$ and resistivities $\rho_q$ ($q=1,...,Q$). Expanding $\rho^{(a)}$ in Taylor series we obtain

$$\Delta \rho^{(a)} = \rho^{(a)} - \rho_0^{(a)} =$$
$$\sum_{q=1}^{Q} \frac{\partial \rho_0^{(a)}}{\partial \rho_{0,q}} \Delta \rho_q + \Sigma(\text{higher-order derivatives}), \tag{5}$$

where $\Delta \rho^{(a)}$ represents the departure of $\rho^{(a)}$ from the apparent resistivity tensor $\rho_0^{(a)}$ of a reference resistivity model which we indicate with $mod_0$. Accordingly, $\Delta \rho_q$ is, in the $q$-th cell, the departure of the actual resistivity $\rho_q$ from the resistivity $\rho_{0,q}$ in $mod_0$.

Since the trace of a sum of matrices is equal to the sum of the traces of the single matrices, using eq.3, we readily obtain

$$\Delta P = P - P_0 =$$
$$\sum_{q=1}^{Q} \frac{\partial P_0}{\partial \rho_{0,q}} \Delta \rho_q + \Sigma(\text{higher-order derivatives}), \tag{6}$$

where $\Delta P$ represents the tensor invariant departure of the actual tensor invariant $P$ from the tensor invariant $P_0$ related to $mod_0$. The term $\partial P_0 / \partial \rho_{0,q}$ is computed using eq.3 as follows

$$\frac{\partial P_0}{\partial \rho_{0,q}} = \tag{7}$$

$$\frac{\left(\frac{\partial E_{0,1x}}{\partial \rho_{0,q}}\right)J_{2y} - \left(\frac{\partial E_{0,2x}}{\partial \rho_{0,q}}\right)J_{1y} + \left(\frac{\partial E_{0,2y}}{\partial \rho0,q}\right)J_{1x} - \left(\frac{\partial E_{0,1y}}{\partial \rho_{0,q}}\right)J_{2x}}{2(J_{1x}J_{2y} - J_{2x}J_{1y})}.$$

If, for simplicity and without loss of generality, we assume that $mod_0$ is a uniform and isotropic half-space, $\partial P_0 / \partial \rho_{0,q}$ can be derived analytically using the Frechet derivative of the electric potential for the uniform half-space (Park and Van, 1991; Loke and Barker, 1995). In fact, the variation of the electrical potential $\phi_i$ at a point $\mathbf{r} \equiv (x,y,0)$ on the earth's surface, due to a small variation of the resistivity in a volume $\Delta V$ immersed in a uniform half-space about the point $\mathbf{r}_q \equiv (x_q, y_q, z_q)$, is

$$\frac{\partial \phi_i}{\partial \rho_{0,q}} = \frac{\partial \phi_{A_i}}{\partial \rho_{0,q}} + \frac{\partial \phi_{B_i}}{\partial \rho_{0,q}}, \tag{8}$$

where

$$\frac{\partial \phi_{A_i}}{\partial \rho_{0,q}} = \frac{I_i}{4\pi^2} \frac{(\mathbf{r}_{A_i} - \mathbf{r}_q) \cdot (\mathbf{r} - \mathbf{r}_q)}{|\mathbf{r}_{A_i} - \mathbf{r}_q|^3 |\mathbf{r} - \mathbf{r}_q|^3}, \quad (9a)$$

$$\frac{\partial \phi_{B_i}}{\partial \rho_{0,q}} = \frac{I_i}{4\pi^2} \frac{(\mathbf{r}_{B_i} - \mathbf{r}_q) \cdot (\mathbf{r} - \mathbf{r}_q)}{|\mathbf{r}_{B_i} - \mathbf{r}_q|^3 |\mathbf{r} - \mathbf{r}_q|^3}. \quad (9b)$$

Omitting the simple but lengthy mathematical steps, from eq.9a and eq.9b we can at last compute the Frechet derivatives of the electrical field components using the expressions

$$\frac{\partial E_{0,ix}}{\partial \rho_{0,q}} = \frac{\partial}{\partial x}\left(\frac{\partial \phi_{A_i}}{\partial \rho_{0,q}} + \frac{\partial \phi_{B_i}}{\partial \rho_{0,q}}\right), \quad (10a)$$

$$\frac{\partial E_{0,iy}}{\partial \rho_{0,q}} = \frac{\partial}{\partial y}\left(\frac{\partial \phi_{A_i}}{\partial \rho_{0,q}} + \frac{\partial \phi_{B_i}}{\partial \rho_{0,q}}\right). \quad (10b)$$

## RESISTIVITY ANOMALY PROBABILITY TOMOGRAPHY

### *The resistivity anomaly occurrence probability*

In order to develop the resistivity anomaly probability tomography (RAPT) method we start by introducing the concept of $\Delta P$-signal energy $\eta$ over the whole survey surface $S$ as

$$\eta = \int_S \Delta P^2(\mathbf{r}) dS. \quad (11)$$

Using the expansion at the right-hand side of eq.6, we extract the main contribution $\eta_1$ related to the first order derivatives as follows

$$\eta_1 = \sum_{q=1}^{Q} \Delta \rho_q \int_S \Delta P(\mathbf{r}) \Im(\mathbf{r} - \mathbf{r}_q) dS, \quad (12)$$

where we have put $\partial P_0 / \partial \rho_{0,q} = \Im(\mathbf{r} - \mathbf{r}_q)$.

Taking the single $q$-th term from eq.12 and applying Schwarz inequality, we obtain

$$\left[\int_S \Delta P(\mathbf{r}) \Im(\mathbf{r} - \mathbf{r}_q) dS\right]^2 \leq \int_S \Delta P^2(\mathbf{r}) dS \cdot \int_S \Im^2(\mathbf{r} - \mathbf{r}_q) dS. \quad (13)$$

Dividing the square root of the left-hand term of the inequality 13 by the square root of the right-hand term,

we can at last introduce a resistivity anomaly occurrence probability (RAOP) function $\eta(\mathbf{r}_q)$ as follows

$$\eta(\mathbf{r}_q) = C_q \int_S \Delta P(\mathbf{r}) \Im(\mathbf{r} - \mathbf{r}_q) dS, \quad (14)$$

where it is

$$C_q = \left[\int_S \Delta P^2(\mathbf{r}) dS \cdot \int_S \Im^2(\mathbf{r} - \mathbf{r}_q) dS\right]^{-1/2}. \quad (15)$$

The RAOP function satisfies the condition

$$-1 \leq \eta(\mathbf{r}_q) \leq +1. \quad (16)$$

Each value of $\eta(\mathbf{r}_q)$ is interpreted as the probability that a resistivity anomaly can obtain in the $q$-th cell, as responsible of the shape pattern of the $\Delta P$-function over $S$. Positive $\eta(\mathbf{r}_q)$ values are associated with increments of resistivity with respect to $mod_0$, while negative values are associated with decrements of resistivity.

The role of probability given to $\eta(\mathbf{r}_q)$ is motivated as follows. As is well known, a probability measure **P** is defined as a function assigning to every subset $\gamma$ of a space of states $\Gamma$ a real number **P**($\gamma$) such that (Gnedenko, 1979)

**P**($\gamma$)≥0, for every $\gamma$, (17a)
**P**($\Gamma$)=1, (17b)

if $\gamma=\alpha\cup\beta$, with $\alpha\cap\beta\equiv 0$,
**P**($\gamma$)=**P**($\alpha\cup\beta$)=**P**($\alpha$)+**P**($\beta$). (17c)

Assuming that the presence of a resistivity departure at $\mathbf{r}_q$ does not depend on the presence of a resistivity departure at another point, the function

$$\mathbf{P}(\mathbf{r}_q) = \frac{|\eta(\mathbf{r}_q)|}{\int_V |\eta(\mathbf{r}_q)| dV}, \quad (18)$$

where $V$ is a generic volume including all non-vanishing values of $|\eta(\mathbf{r}_q)|$, can be defined as a probability density, allowing a measure of the probability to get a resistivity departure at $\mathbf{r}_q$ to be obtained in agreement with axioms (17a,b,c).

Actually, the definition given in eq.14 differs from that in eq.18 for an unknown constant factor appearing at the denominator of eq.18, and has the advantage of giving information on the sign of the sources. Therefore, $\eta(\mathbf{r}_q)$ can conventionally be assumed as a measure of the resistivity anomaly occurrence probability.

*The resistivity anomaly probability tomography*

The 3D RAPT imaging approach consists in a cross-correlation procedure performed by the scanner function $\Im(\mathbf{r}-\mathbf{r}_q)$ over the data function $\Delta P(\mathbf{r})$ within a volume, called the tomospace, lying below the survey area *S*. In practice, we utilise an elementary cell with a positive resistivity anomaly of unitary strength to scan the whole tomospace and search where resistivity variations with respect to a reference $mod_0$ are placed in a probabilistic sense. For each position of the scanning element, i.e. for each value of *q*, the corresponding value of $\eta(\mathbf{r}_q)$ is calculated using a discretized version of eq.14 given as

$$\eta_q = C_q \sum_S \Delta P \Im_q , \qquad (19)$$

with

$$C_q = \left( \sum_S \Delta P^2 \cdot \sum_S \Im_q^2 \right)^{-1/2} . \qquad (20)$$

Each value of $\eta_q$ is attributed to the central point of the scanning elementary cell. The final step is a regular grid of values of $\eta_q$, which can be contoured in order to obtain a tomography imaging of the scanned volume.

**A SYNTHETIC EXAMPLE**

To test the resolving power of the RAPT method, we consider the synthetic example of a composite prismatic target immersed in a uniform half-space. Fig.1 shows a plan and section view of a two-block structure, where a resistive prism is coupled with a conductive composite prism. This synthetic example has been chosen in order to analyze the response of the method on a model best approximating the geometry of a known archaeological structure which will be discussed later as field example. In fig.1, $A_1B_1$ and $A_2B_2$ are a pair of orthogonal current bipoles used to simulate the tensorial resistivity survey. Each current bipole is assumed to be 19 m long, and a passive 0.5 m long dipole is supposed to be moved at steps of 0.5 m along crossed profiles spaced 0.5 m apart, within an area of 9×9 m². A finite element program has been edited to obtain the synthetic apparent resistivity response due to the model. Then, a total of 361 apparent resistivity data has been processed to obtain the RAPT simulation.

The top slice in Fig.2 shows the anomaly map of *P*, corresponding to the resistivity model of Fig.1. The data are expressed in Ωm and the contour scale is reported on top. The following slices in the same figure show the RAPT images at various depths. The hosting half-space with resistivity 100 Ωm has been assumed as $mod_0$. The horizontal slices are drawn every 0.5 m of depth beneath the survey surface. The RAOP contour scale is drawn at the bottom. The RAPT images show a positive RAOP nucleus exactly in correspondence with the position of the resistive body. The highest $\eta_q$ values occur at 2.5 m below the survey plane. The slices show also a negative RAOP nucleus extending downward exactly below the conductive composite body. The lowest $\eta_q$ values occur at 1.5 m beneath the survey surface.

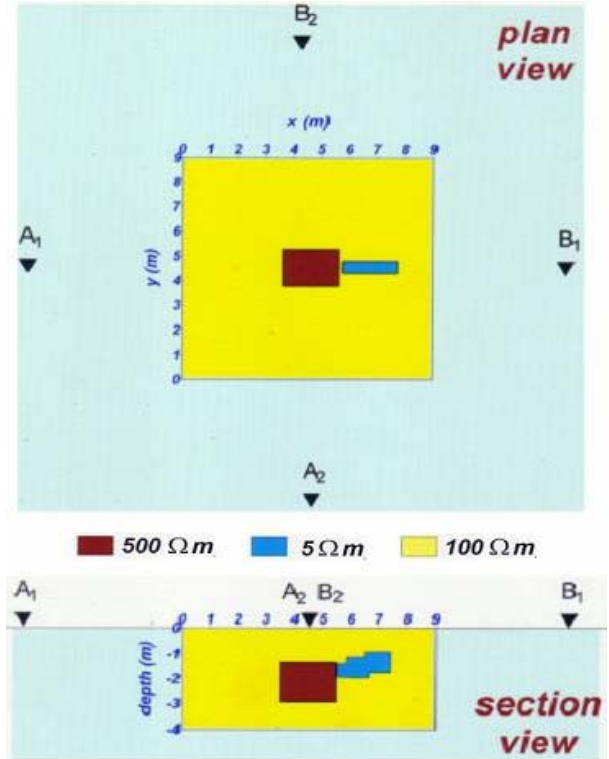

**Figure 1** A synthetic example: the two-block model immersed in a uniform half-space. $A_1B_1$ and $A_2B_2$ represent the orthogonal energizing current bipoles.

This example shows that the RAPT method is quite able to distinguish two adjacent bodies with contrasting resistivity, and to locate the equivalent physical sources of the anomalies at a position underground, nearly corresponding with the barycentre of the bodies. To better appreciate the resolving power of the new method, fig.3 shows, for comparison, the tomographic slices obtained by considering only the central resistive prism. It can be observed that, except for the values of $\eta_q$ which are now greater than those in fig.2 at the same points, the pattern of the whole $\eta_q$ representation is quite similar to that in fig.2. This means that if the target had been the central prism, its identification pattern would have not been dramatically distorted by the presence of the lateral disturbing prism with contrasting resistivity.

# A FIELD EXAMPLE

*Site description*

In the Sabine necropolis at Colle del Forno, located in the Tiber valley, 30 km north of Rome, Italy, a site not yet exploited was chosen to test the applicability of the new 3D tomography method in an actual archaeological context. The same area was previously explored to test a former geoelectrical tomographic method based on the concept of charge occurrence probability (Mauriello *et al*., 1998). The existence of hypogeal *dromos*-chamber tombs was strongly supported by integrated geophysical surveys, including the self-potential, dipole geoelectrics, ground penetrating radar and differential magnetometry methods (Cammarano *et al*., 1997 a,b; 1998).

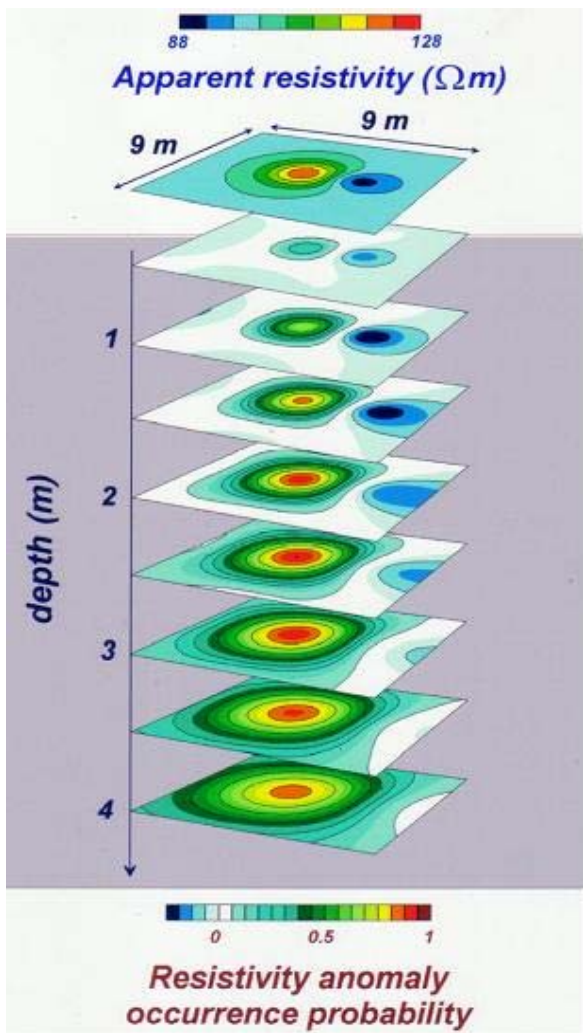

**Figure 2** Simulated resistivity source element 3D probability tomography for the synthetic model of fig.1. The slice at the top is the synthetic survey map of the trace of the apparent resistivity tensor.

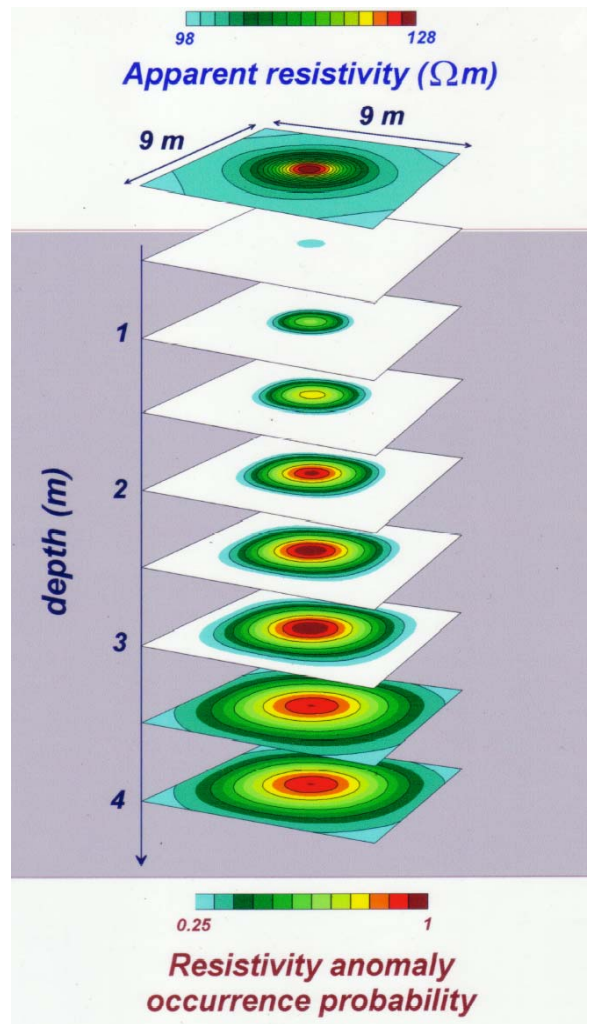

**Figure 3** Simulated resistivity source element 3D probability tomography for a single prism model, corresponding with the red block in fig.1. The slice at the top is the synthetic survey map of the trace of the apparent resistivity tensor.

Fig.4 displays sketched plan and section views of a standard tomb in the Sabine necropolis (Santoro, 1977). It consists of two distinct volumes. The main body is the tomb chamber with a standard volume of 2×2×2 m$^3$. Its roof is normally found at an average depth of 1 m b.g.l.. The accessory structure is a downward sloping corridor (*dromos*), up to 6 m long and with a mean 1×1 m$^2$ cross-section. The tombs were excavated in a uniform layer of lithoid tuff with a mean thickness of 10 m, characterized by a resistivity value in the range 20-30 Ωm. The tuff layer overlies a thick Pleistocene-Quaternary sands and clays alternate sequence, and is covered by a 20-30 cm thick clayey-sandy top soil. The tomb chambers have generally been found in a good state of conservation, sometimes partially filled with loose sediments, while the *dromos* have almost always been found completely filled with wet loose sediments, showing a resistivity of the order of 10 Ωm or less.

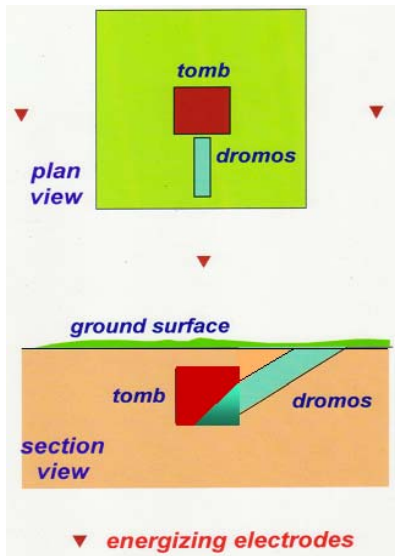

**Figure 4** A field example. Plan and section views of a typical dromos-chamber tomb in the Sabine necropolis at Colle del Forno near Rome, Italy.

In synthesis, we are faced with a two-body problem, featured, at least in principle, as a highly resistive body (the totally or partially void tomb chamber) joined with a conductive body (the completely filled *dromos*), both enclosed in a nearly uniform medium with intermediate resistivity. A close similarity can be observed between the model of fig.4 and the synthetic model of fig.1.

*Field technique*

We employed a low-frequency AC energizing unit. A current of amplitude 100 mA and frequency 128 Hz was injected into the ground, and the potential drops $\Delta\phi_i$ at the same frequency were measured across a 0.5 m long dipole. The apparent resistivities were computed using the standard formula for DC geoelectrics, since the ratio of the survey probing length (14 m at most) to the current wavelength (140 m at least, in a minimum 10 Ωm resistivity environment) was quite negligible, of the order of 10%, at most. In other words, we admit that in the quasi-static limit in which we carried out the field experiment, the current density **J** can be considered a divergence-free vector as in DC geoelectrics.

The first $A_1B_1$ current layout for the experimental determination of the apparent resistivity tensor was a 19 m long bipole spread along the N-S median axis of the 9×9 m² survey area, symmetrically on one and the other side (see top slice in fig.5). The $M_1N_1$ potential dipole was first moved with a sampling step of 0.5 m along parallel straight profiles in the N-S direction, spaced 0.5 m apart. Then, a second set of data was collected with the $M_1N_1$ dipole moved along parallel straight profiles in the E-W direction, again spaced 0.5 m apart. The two sets allowed the electric field vector $\mathbf{E}_1$ to be determined at the nodes of the profile grid. Accordingly, the second $A_2B_2$ current layout was an E-W 19 m long bipole, again crossing symmetrically the area. The same procedure as before was utilised to collect at the same points of the profile grid the data useful for the determination of the electric field vector $\mathbf{E}_2$.

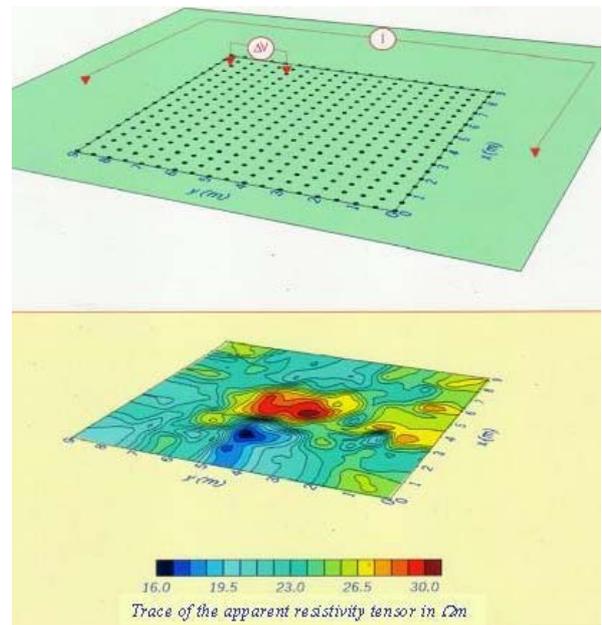

**Figure 5** A field example. Top slice: the electrode array in the survey area. Bottom slice: trace of the apparent resistivity tensor field map above the surveyed dromos-chamber tomb in the Sabine necropolis at Colle del Forno near Rome, Italy.

**Results and discussion**

The anomaly map of the rotational invariant *P* obtained by this procedure is drawn in the bottom slice of Fig.5. A couple of anomalies with opposite trend appears very well defined in the central-western sector of the survey area. In total agreement with the results from previous surveys (Cammarano et al., 1997, 1998), the large high of the *P* invariant (red anomaly) is ascribed to the main chamber of the Sabine tomb. Accordingly, the narrow low of the *P* invariant (blue anomaly) is associated with the entrance corridor (*dromos*), completely filled with loose sediments.

Before illustrating the RAPT imaging, it may result quite interesting to show the greater performance of the apparent resistivity tensor method compared with that of the traditional scalar apparent resistivity method.

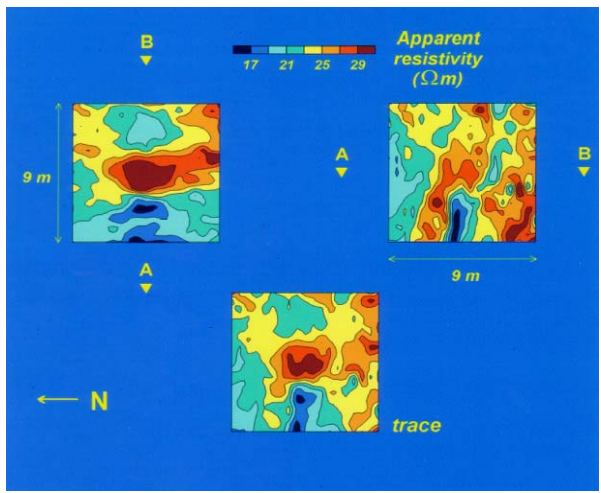

**Figure 6** A field example. The apparent resistivity tensor map above the dromos-chamber tomb in the Sabine necropolis at Colle del Forno, Rome, Italy (bottom slice), compared with the scalar apparent resistivity maps for the E-W (left top slice) and N-S (right top slice) polarizations of the AB current source layout.

The distortion due to the polarization of each single AB electric current line is quite evident. A deflation and an inflation of the anomalies occur respectively parallel and perpendicular to the main direction of the primary current flow. On the contrary, the combined use of the two AB polarization for the determination of the trace of the apparent resistivity tensor provides a completely distortion-free map with more confined anomalies over the source bodies.

For the tomographic elaboration we have assumed as $mod_0$ a uniform half-space with a resistivity of 24 Ωm, equal to the average resistivity of the lithoid tuff about the current electrodes. In Fig.7, the sequence of slices below the anomaly map show the RAOP tomography every 0.25 m from the ground level down to 3.5 m of depth. Only values of $\eta(\mathbf{r}_q)$ exceeding in modulus 0.3 have been reported. The first evidence is the positive RAOP nucleus in the central part, which unequivocally indicates the existence of the empty tomb chamber with the position of its barycentre placed, as expected, within the depth range 1.5-2 m b.g.l., where the highest $\eta(\mathbf{r}_q)$ values occur. The second signal is the negative nucleus in the central-western side of the slices, which would thus highlight the entrance corridor to the tomb, filled with conductive sandy-clayey sediments. However, the lowest $\eta(\mathbf{r}_q)$ values appear concentrated around 2.5 m of depth, well beyond the expected depth of the barycentre of the *dromos* structure. Very likely, the minimum at 2.5 m of depth would most properly indicate the area where resistivity reaches its lowest value inside the corridor, due to an increased accumulation of water in the basal clayey-sandy deposit.

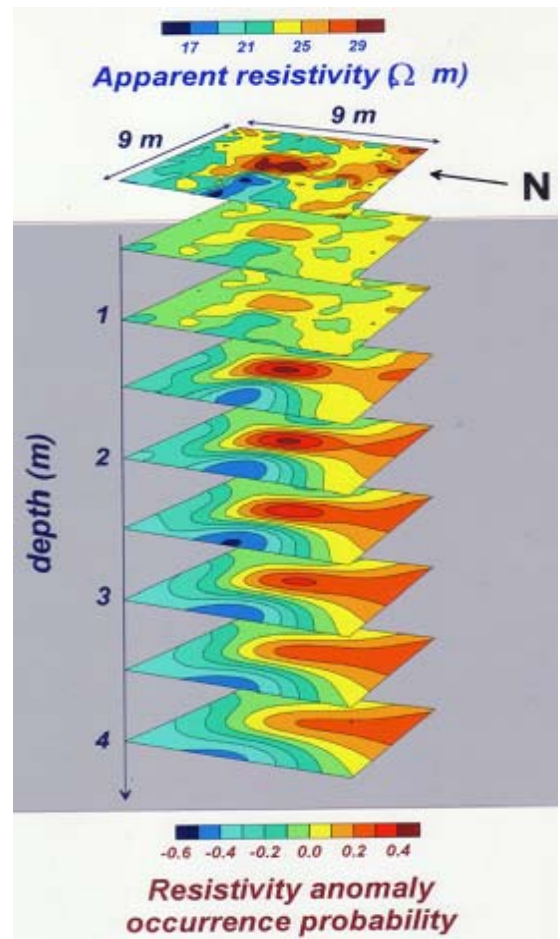

**Figure 7** Field example. The resistivity source 3D probability tomography for the field case of the Sabine tomb. The slice at the top is the experimental survey map of the trace of the apparent resistivity tensor.

## CONCLUSION

The 3D probability tomography previously developed for the scalar resistivity method has been here extended to the 2D tensorial apparent resistivity acquisition mode. It has been demonstrated that the trace of the apparent resistivity tensor provides distortion-free maps with the anomalies closely confined on the source bodies. Such a property strongly enhance the performance of the high-resolution, target-oriented probability tomography that has been proposed in this paper.

The new tomography method has the advantage that, in principle, no strict reference to the geometry of the sources of anomaly is needed as an a priori constraint to start with the imaging algorithm. It only relates to the pure physical aspects of the electrical stimulation of the buried structures. The use of a probability parameter for resistivity pattern recognition underground is thought to be unavoidable. Indeed, due to intrinsic equivalence and cultural and/or natural noise contamination sources, the search for a deterministic solution of the true shape and size of target bodies has basically much less common-sense than it is believed.